\documentstyle[12pt]{article}
\font\frak=eufm10 scaled\magstep1

\font\black=msbm10 scaled\magstep1
\font\bigblack=msbm10 scaled\magstep 2
\font\bbigblack=msbm10 scaled\magstep3

\def\goth #1{\hbox{{\frak #1}}}

\def\field #1{\hbox{{\black #1}}}
\def\bigfield #1{\hbox{{\bigblack #1}}}
\def\bbigfield #1{\hbox{{\bbigblack #1}}}
\def\Bbb #1{\hbox{{\black #1}}}
\def\v #1{\vert #1\vert}             
 
\def\m #1 #2{(-1)^{{\v #1} {\v #2}}} 
\def\pd#1#2{\frac{\partial#1}{\partial#2}}
\def\<#1>{\langle#1\rangle}        
\def\>#1{{\bf #1}}                
\def\f(#1,#2){\frac{#1}{#2}}

\def\dt2#1{\frac{d^2 #1}{dt^2}}
\def\matriz#1#2{\left( \begin{array}{#1} #2 \end{array}\right) }
\def\Eq#1{{\begin{equation} #1 \end{equation}}}
\def\deter#1#2{\left| \begin{array}{#1} #2 \end{array}\right| }

\def\a{\alpha}
\def\G{{\Gamma}}
\def\la{\lambda}                   
\def\w{\omega}                     
	     
\def\R{{\hbox{{\field R}}}}             
\def\big R{{\hbox{{\bigfield R}}}}
\def\bbig R{{\hbox{{\bbigfield R}}}}

\def\ad{{\hbox{ad}}}

\def\Arg{\hbox{{\rm Arg\,}}}

\def\be{\begin{equation}}
\def\ee{\end{equation}}
\def\bea{\begin{eqnarray}}
\def\eea{\end{eqnarray}}

\begin{document}

\centerline{\Large\bf The non-linear superposition principle} 

\bigskip

\centerline{\Large\bf and the Wei--Norman method} 

\vskip 2cm

\centerline{ {\sc J.F. Cari\~nena*,  G. Marmo${}^{\mbox{\small{\S}}}$ 
  and  J. Nasarre${}^{\mbox{\small{\S\S}}}$ 
}}

\vskip 1cm

\centerline{*Depto. F\'{\i}sica Te\'orica, Universidad de Zaragoza, 50009 
Zaragoza,
Spain.}

\centerline{${}^{\mbox{\small{\S}}}$Dipto. di Scienze Fisiche, Universit\'a di Napoli,
80125 Napoli, Italy.}

 \centerline{${}^{\mbox{\small{\S\S}}}$Seminario~de Matem\'aticas, IES Miguel Catal\'an,
 50009 Zaragoza,  Spain. }

\vskip 2cm

\centerline{\sc Abstract}
{\small Group theoretical methods are used to study some properties
of the Riccati equation, which is the only differential equation
admitting a nonlinear superposition principle. The Wei--Norman method
is applied for obtaining
the associated differential equation in the group $SL(2,\R)$.
The superposition principle for first order differential
equation systems and Lie-Scheffers theorem are also analysed from
this group theoretical perspective. Finally, the theory is  applied
in the solution of second order
differential equations like time independent Schr\"odinger equation.
 }

\vskip 1cm

\def\PRL{Phys. Rev. Lett.}  
\def\NPB{Nucl. Phys.}  
\def\CMP{Commun. Math. Phys.}  
\def\PRD{Phys. Rev.} 
\def\PLA{Phys. Lett.} 
\def\PLB{Phys. Lett.} 
\def\JMP{J. Math. Phys.} 
\def\PTP{Prog. Theor. Phys.} 
\def\SPTP{Suppl. Prog. Theor. Phys.} 
\def\AP{Ann. of Phys.} 
\def\PNAS{Proc. Natl. Acad. Sci. USA} 
\def\RMP{Rev. Mod. Phys.}  
\def\PR{Phys. Reports} 
\def\AM{Ann. of Math.} 
\def\FAA{Functional Analysis and Its Application} 
\def\BAMS{Bull. Am. Math. Soc.}  
\def\TAMS{Trans. Am. Math. Soc.}  
\def\InvM{Invent. Math.}  
\def\LMP{Letters in Math. Phys.} 
\def\IJMPA{Int. J. Mod. Phys.}  
\def\AdM{Advances in Math.}  
\def\RMaP{Reports on Math. Phys.}  
\def\APP{Acta Phys. Polon.}  
\def\TMP{Theor. Mat. Phys.}  
\def\JPA{J. Physics} 
\def\MPLA{Mod. Phys. Lett.} 
\def\JETP{Sov. Phys. JETP} 
\def\JETPL{ Sov. Phys. JETP Lett.} 
\def\PHS{Physica}

\section{
Introduction}

Nonlinear phenomena are having everyday more and more importance
in almost all branches of science,
and in particular in Physics. One of the most important nonlinear
differential equation 
is the  Riccati equation (see e.g. \cite{HTD} and references therein).
This differential equation has recently been shown to be related
with the factorization method
(see e.g.\cite{{ASt},{ASt2},{SR},{RS}}). The recent interest of this  equation is steadly
increasing since Witten's introduction of supersymmetric
Quantum Mechanics \cite{{Wi}, {LRB}}.

Two important features of this Riccati type differential equation are:

i)  Even if we cannot find the general solution by means of a finite
number of quadratures over elementary functions of the coefficients
$a_i(t)$ defining the differential equation,
\be
\frac{dx(t)}{dt}=a_0(t)+a_1(t)\,x(t)+a_2(t)\,x^2(t),
\label{Riceq}
\ee
once a particular solution $x_1(t)$ is known, the change of variable $x=x_1+z$ leads to a new
differential equation of the Bernouilli (for $n=2$) type:
\be
\frac {dz}{dt}=(2\, a_2\, x_1+a_1) z+a_2\,z^2. \label{Bereq}
\ee

This is a particular instance of (\ref{Riceq}) for which $a_0=0$.
Notice that under  the change of variable $w= 1/x$  the Riccati  equation
(\ref{Riceq}) becomes
\be
\frac{dw(t)}{dt}+a_0(t)\, w^2(t)+a_1(t)\,w(t)+a_2(t)=0.
\ee

In particular, the Bernouilli equation (\ref{Bereq})
 can be reduced to a linear one just by introducing the
new variable $u=1/z$. In this way, if we know a particular solution
the general solution can be found through two quadratures.

ii) When three particular solutions, $x_1(t),x_2(t),x_3(t)$, of
the differential equation (\ref{Riceq}) are known,
the general solution can be written with no quadrature at all:
\be
\frac{x-x_1}{x-x_2}:\frac{x_3-x_1}{x_3-x_2}=k ,
\ee
where $k$ is an arbitrary constant determining each particular solution.

In this sense we can say that there exists a nonlinear superposition
principle for the Riccati equation, because the general solution can be
expressed as a function $x=\Phi (x_1,x_2,x_3,k)$ of three particular
solutions and one arbitrary constant $k$.

Our aim is to explain these facts from a group theoretic viewpoint and
present some new ideas both about the Riccati equation itself and on the
nonlinear superposition principle, for which the Riccati equation is
the simplest case.

The organization of this paper is as follows.    In Section 2 we review 
the problem of reducing 
a second order linear differential equation to a nonlinear first order  
Riccati equation, what means that the original linear superposition 
principle for  the second order equation should be replaced by a 
nonlinear superposition principle. We also remark that this fact is due to 
a relation of the Riccati equation with the $SL(2,\R)$ group
to be explicited later. 
In Section 3 we explain a method developed by Wei and Norman \cite{WN}
for determining 
the solution of a differential equation in a Lie group and we apply 
the method for the study of the Riccati equation, finding in this way 
the explicit form of the superposition principle as a consequence of
some group theoretical computations.   The superposition principle for
first order
differential equation systems and Lie--Scheffers theorem are studied in
Section 4. It is
shown that for an important class of such systems the problem of finding
 the general solution is
reduced to the simpler problem of finding one particular solution of
another system  on a Lie group $G$,
and moreover, even without solving directly this new system the solution
of the original
system can be easily
found as soon as we know a fundamental set of solutions of it. Once again the simplest
case is Riccati equation and the superposition principle can be found by
determining the first integral of a system. Finally in Section 5 we give as an example the application
of the Wei--Norman method in the solution of second order differential equations
taking as a propotype the Schr\"odinger equation for the harmonic
oscillator.

\section{The nonlinear superposition principle}
\setcounter{equation}{0}

There is a well known method of relating a linear
second order differential equation
with a Riccati equation. Actually, given the linear second order differential equation
\be
\frac {d^2u}{dt^2}+b(t)\,\frac {du}{dt}+c(t)\,u=0,\label{LSODE}
\ee
the property of linearity means that the vector field
$$X=u\,\pd{}{u}$$
generates a one--parameter
Lie group of point symmetries of the equation (\ref{LSODE})
$$\bar t(\epsilon)=t,\quad  \bar u(\epsilon)=e^\epsilon u.
$$

Changing coordinate $u$ to a new  one $v=\varphi (u)$ such that 
the vector field  $X=u\partial/\partial u$ becomes
a translation generator
(Straightening--out Theorem), i.e. $X=\partial/\partial v$, and
therefore determined by the equation
$Xv=1$, leads to $v=\log u$, i.e. $u=e^v$, and then,
\be
\frac {du}{dt}=e^v \, \frac {dv}{dt}=u\, \frac {dv}{dt}.\label{changev}
\ee
When written in
terms of this new coordinate
the equation (\ref{LSODE}) becomes 
$$
\frac {d^2v}{dt^2}+b(t)\, \frac {dv}{dt}+\left(\frac {dv}{dt}\right)^2+c(t)=0.
$$

The unknown function $v$ does not appear in the preceding
equation and therefore a lowering
 of the order is obtained when introducing the change of variable
$x=\frac {dv}{dt}$, and then we will get a Riccati equation
\be
\frac {dx}{dt}=-c-bx-x^2,\label{asocR}
\ee
as it was pointed out in  \cite{DRS}.

 Therefore, the linear superposition principle for solutions
of (\ref{LSODE}) translates in
 a nonlinear superposition principle for those of this Riccati
equation, as it
will be shown later.

 Notice that (\ref{changev}) shows that
 \Eq{x=\frac 1u\, \frac{du}{dt}.\label {varred}}

 The second order differential equation (\ref{LSODE}) is equivalent to the set of 
 (\ref{asocR})  and (\ref{varred}).
 We should also remark that two solutions $u_1,u_2$ of (\ref{LSODE}) project on the
 same solution of (\ref{asocR}) if and only if there exists a nonzero real
number $\lambda\in \R$ such that $u_2(0)=\lambda u_1(0)$ and
$u'_2(0)=\lambda u'_1(0)$.

 From the geometric viewpoint, Riccati equation can be interpreted as the
one determining the integral curves of a time--dependent  vector field
$$\Gamma=(a_0(t)+a_1(t)x+a_2(t)x^2)\pd{}{x}.\label{vfRic}
$$

Let us remark that this vector field can be written as a linear
combination with time dependent coefficients of the vector fields
\be
L_0 =\pd{}{x}, \quad
L_1 =x\,\pd{}{x} ,\quad
L_2 = x^2\,\pd{}{x},\label{sl2gen}
\ee
which generate a three dimensional real Lie algebra with defining
relations
\be
[L_0,L_1] =L_0,\quad 
 [L_0,L_2] =  2L_1 \quad
[L_1,L_2]  = L_2  ,
\ee
and therefore isomorphic to  $\goth{sl}(2,\R)$. In fact, it is an easy matter
to check that the (local) one--parameter transformation Lie groups
of $\R$ generated by $L_0$,
$L_1$ and $L_2$ are
$$ x\mapsto x+\epsilon,\quad x\mapsto e^\epsilon x,\quad x
\mapsto\frac x{1-x\epsilon },
$$
i.e., they are  fundamental vector fields corresponding
to the action of
$SL(2,\R)$ on the real line $ {\R}$ extended with a point at the infinity,
$\bar {\R}$, given by
$$\Phi(A, x)=\frac {\alpha x+\beta}{\gamma x+\delta},
\qquad A=\matriz{cc}{\alpha &\beta\\
 \gamma &\delta},\ {\rm if}\ x\not =-\frac\delta\gamma,
$$
and
$$\Phi(A,\infty)=\frac \alpha\gamma,\quad \Phi(A,-\frac\delta\gamma)=\infty.
$$

\section{The Wei--Norman method}
\setcounter{equation}{0}

Let us consider a differential equation system
\be
\frac{dx^i(t)}{dt}=X^i(x,t) ,\qquad i=1,\ldots,n,
\ee
which can be seen as the differential equation system whose solutions are
the integral curves of the time--dependent vector field
\bea X=X^i(x,t)\, \pd {}{x^i}.
\eea

The theorem for existence and uniqueness of solutions of the preceding
differential equation tells us that, for small enough $t$,
 there exists a map $\Phi_t$ applying
the initial condition $x(0)=(x^i(0)) $ into the corresponding
value $x^i(t)$. Correspondingly, functions $f$ defined in a neigborhood
of $x(0)$  transform as 
\be
[U(t)f](x)=f\left(\Phi_t^{-1}(x)\right),
\ee
and taking derivatives with respect to $t$ we obtain
\begin{eqnarray}
\left[\frac{dU(t)}{dt}f\right](x)&=&\frac{d}{dt}\left((f\circ \Phi_t^{-1})
(x)\right)=\frac {dx^i}{dt}\pd f{x^i}(\Phi_t^{-1}(x))\cr
&=&X(f)\left(\Phi_t^{-1}(x)\right)=[U(t)(Xf)](x).
\end{eqnarray}

This relation is valid for any differentiable function $f$,
and therefore
$$
\frac{dU(t)}{dt}=U(t)X.
$$

We recall \cite{WN}   that  given such a differential equation for
operators for $X$ being a linear combination of vector fields in $\R^n$, 
$$X=\sum_{k=1}^ma_k(t)L_k,
$$
namely,
\be
\frac{dU(t)}{dt}= \sum_{k=1}^ma_k(t)U(t)L_k,
\ee
where the  $L_k$ span a finite dimensional real Lie  algebra, it is possible to write  
the general solution   in the form 
\be
U(t)=\prod_{i=1}^{m}\exp(g_i(t)L_i),
\ee
where the functions $g_i(t)$   are given by the solution of the system
obtained from the relation
\be
\sum_{i=1}^{m}a_i(t)L_i=\sum_{i=1}^{m}\dot{g}_i(t)
\left[\prod_{j=i+1}^{m}\exp(-g_j(t)\ad{L_j})\right]L_i,\label{WNrel}
\ee
and the initial condition $g_i(0)=0,\ i=1,\cdots,m$.

This method proposed by Wei--Norman (see also \cite{DGT})    can be used in the case of the 
 Riccati  equation and the generalization for other differential equation system involving
several degrees of freedom is immediate. In fact, given the differential equation (\ref{Riceq}) 
there will be an evolution operator $U(t)$ which takes values in $SL(2,\R)$ 
and satisfies the differential equation
\be
\frac{dU(t)}{dt}=U(t)[a_0(t)L_0+a_1(t)L_1+a_2(t)L_2],
\ee
together with the initial condition $U(0)=I$, 
and  where the vector fields  $L_k$, for $k=0,1,2$, 
are  fundamental vector fields   associated to 
the left action of $SL(2,\R)$ on the extended real line   $\bar\R$ that are 
explicitly given in (\ref{sl2gen})  and generate
a Lie algebra isomorphic to $\goth{sl}(2,\R)$.

According to the  Wei--Norman method \cite{WN},
   there   will be functions $u_0(t),u_1(t),u_2(t)$
such that 
$u_0(0)=u_1(0)=u_2(0)=0$  and 
\be
U(t)=\exp(u_1L_1)\exp(u_2L_2)\exp(u_0L_0).\label{evolop}
\ee

Then, when using (\ref{WNrel}), we obtain 
\begin{eqnarray}
a_0(t)L_0+a_1(t)L_1+a_2(t)L_2&=&
\dot{u}_1(t)\exp(-u_0(t)\ad{L_0})\exp(-u_2(t)\ad{L_2})L_1\nonumber\\&+&
\dot{u}_2(t)\exp(-u_0(t)\ad{L_0})L_2 +\dot{u}_0(t)L_0 \nonumber\\
&=&\dot{u}_1(L_1-u_0L_0+u_2L_2-2u_0u_2L_1+u_0^2u_2L_0)\nonumber \\
&+&\dot{u}_2(L_2-2u_0L_1+u_0^2L_0)+\dot{u}_0L_0,\nonumber
\end{eqnarray}
and therefore the system of differential equations for the functions $u_i$ 
\begin{eqnarray}
a_0(t)&=&-u_0\dot{u}_1+u_0^2u_2\dot{u}_1+u_0^2\dot{u}_2+\dot{u}_0 \cr
a_1(t)&=&\dot{u}_1-2u_0u_2\dot{u}_1-2u_0\dot{u}_2 \cr
a_2(t)&=&u_2\dot{u}_1+\dot{u}_2,
\end{eqnarray}
that can be written in normal form,
\begin{eqnarray}
\dot{u}_0(t)&=&a_0(t)+a_1(t)u_0(t)+a_2(t)u_0^2(t) \cr
\dot{u}_1(t)&=&a_1(t)+2a_2(t)u_0(t)   \cr
\dot{u}_2(t)&=&a_2(t)-a_1(t)u_2(t)-2a_2(t)u_0(t)u_2(t).\label{eqcc}
\end{eqnarray}  

We remark that the first equation for $u_0$ is nothing but the original Riccati equation
and therefore it seems that there is no advantage at all. However
 when looking in the other
two  equations, we see that provided that the appropriate solution
for the first equation, the one determined by $u_0(0)=0$ has
been found, the solution for the second one is almost immediate and
when introducing these values in the third equation this one becomes a first
order differential equation and the solution reduces to a
quadrature.
In this sense we have reduced the problem of finding the general solution of
(\ref{Riceq}) to the one of finding the particular solution 
such that $u_0(0)=0$.  This is quite similar to the property i) we 
mentioned in the Introduction.

Once the solution of (\ref{eqcc}) determined by $u_0(0)=u_1(0)=u_2(0)=0$
has been found, the general solution of the differential equation
will be written
\be
x(t)=(U(t)x)(0)=[\exp(u_1L_1)\exp(u_2L_2)\exp(u_0L_0)x]{\vert}_{t=0},\label{gral}
\ee
and therefore
\be
x(t)=\frac{e^{u_1}x_0}{1-u_2e^{u_1}x_0}+u_0,\label{solg}
\ee
where $x_0=x(0)$.

Let us now fix three independent initial conditions. A possible set
is the one given by
$x_1(0)\rightarrow \infty$,
$x_2(0)=0$  y  $x_3(0)=1$, (actually any other three different initial conditions
will be transformed to this one under an appropriate transformation of the 
group $SL(2,\R)$). 

Having in mind the form (\ref{solg}) of the general solution, we see that
the above mentioned initial conditions determine the particular solutions
\begin{eqnarray}
x_1(t)&=&-\frac{1}{u_2(t)}+u_0(t) \cr
x_2(t)&=&u_0(t)  \cr
x_3(t)&=&\frac{e^{u_1(t)}}{1-u_2(t)e^{u_1(t)}}+u_0(t) ,
\end{eqnarray}     
and then the functions $u_0,u_1,u_2$ are determined as
\begin{eqnarray}
u_0(t)&=&x_2(t) \\
u_1(t)&=&\log \left[\frac {(x_3(t)-x_2(t))(x_2(t)-x_1(t))}{(x_3(t)-x_1
(t))}\right] \\
u_2(t)&=-&\frac {1}{x_2(t)-x_1(t)}  ,
\end{eqnarray} 
and therefore, when putting these values in (\ref{solg}), we find that
for the general solution
\be
x(t)=\frac {x_0x_1(t)(x_3(t)-x_2(t))+x_2(t)(x_1(t)-x_3(t))}{x_0(x_3(t)-x_2(t))
+(x_1(t)-x_3(t))},\label{solgxp}
\ee
which is the well  known superposition
principle for
Riccati equation (\ref{Riceq}) which can also be written as
\be
\frac{(x-x_2)(x_3-x_1)}{(x-x_1)(x_3-x_2)}=x_0.
 \ee

  The factorization (\ref{evolop}) for the evolution operator is not the only possible one,
  but there are other five alternative factorizations.  We next present the results for all possible reorderings:

  \bigskip
  
II)  When we consider the factorization 
$$
U(t)=\exp(g_0L_0)\exp(g_1L_1)\exp(g_2L_2)  ,
$$
the associated system turns out to be 
\begin{eqnarray}
\dot{g}_0&=&a_0e^{-g_1}\cr
\dot{g}_1&=&a_1-2a_0g_2\cr
\dot{g}_2&=&a_2-a_1g_2+a_0g_2^2,      \label{sistg}
\end{eqnarray}
with $g_0(0)=g_1(0)= g_2(0)= 0$,
and then the solution is  
\be
x(t)=\frac {e^{g_1}(x_0+g_0)} {1-g_2e^{g_1}(x_0+g_0)},
\ee
the three particular solutions
 $x_1, x_2, x_3 $  being expressed as
\begin{eqnarray}
x_1(t)&=& -\frac {1}{g_2} \cr
x_2(t)&=& \frac {g_0e^{g_1}}{1-g_0g_2e^{g_1}} \cr
x_3(t)&=& \frac {e^{g_1}(1+g_0)}{1-g_2e^{g_1}(1+g_0)},
\end{eqnarray}
with the inverse relation 
\begin{eqnarray}
g_0&=&\frac {x_2(x_1-x_3)} {x_1(x_3-x_2)} \cr
g_1&=&\log \left[\frac {x_1^2(x_3-x_2)} {(x_1-x_2)(x_1-x_3)}\right] \cr
g_2&=&-\frac {1} {x_1} .
\end{eqnarray}

We should remark that  the third equation in (\ref{sistg}) is a new Riccati 
equation and that once the solution for this new Riccati equation is found, we 
substitute its value in the second  one and integrate without any difficulty, 
and when this value for $g_1$ is put in the first equation we can integrate it. 
Therefore the new result we have found here is the following:  
if one knows a solution of a related Riccati equation, 
given by the third one in 
(\ref{sistg}) equation,  satisfying $g_2(0)=0$, the general solution  
 of (\ref{Riceq}) can be found in a straightforward way.

 \bigskip

III) The third possibility corresponds to the factorization
\be
U(t)=\exp(h_2L_2)\exp(h_1L_1)\exp(h_0L_0),
\ee
the associated system being
\begin{eqnarray}
\dot{h}_0&=&a_0+a_1h_0+a_2h_0^2\\
\dot{h}_1&=&a_1+2a_2h_0\\
\dot{h}_2&=&a_2e^{h_1}    , \label{sisth}
\end{eqnarray}
and the general solution is then 
\be
x(t)=\frac {e^{h_1}x_0} {1-h_2x_0} +h_0.
\ee
The three particular solutions $x_1, x_2, x_3 $  will be written as 
\begin{eqnarray}
x_1(t)&=&-\frac {e^{h_1}}{h_2}+h_0 \cr
x_2(t)&=& h_0 \cr
x_3(t)&=& \frac {e^{h_1}}{1-h_2}+h_0,
\end{eqnarray}
the inverse relations being
\begin{eqnarray}
h_0&=& x_2 \cr
h_1&=&\log \left[\frac {(x_3-x_2)(x_2-x_1)} {(x_3-x_1)}\right] \cr
h_2&=&\frac {x_2-x_3} {x_1-x_3} .
\end{eqnarray}

Notice that in this approach the first equation in (\ref{sisth}) is the original 
equation (\ref{Riceq}).

\bigskip

IV) The fourth reordering leads to the factorization
\be
U(t)=\exp(f_1L_1)\exp(f_0L_0)\exp(f_2L_2),
\ee
associated system
\begin{eqnarray}
\dot{f}_0&=&a_0+a_1f_0-2a_0f_0f_2\cr
\dot{f}_1&=&a_1-2a_0f_2\cr
\dot{f}_2&=& a_2-a_1f_2+a_0f_2^2  ,      \label{nsistf}
\end{eqnarray}
and general solution
\be
x(t)=\frac {e^{f_1}x_0+f_0} {1-f_2(e^{f_1}x_0+f_0)} .
\ee

The three particular solutions are now
\begin{eqnarray}
x_1(t)&=& -\frac {1}{f_2} \cr
x_2(t)&=&\frac {f_0} {1-f_2f_0} \cr
x_3(t)&=& \frac {e^{f_1}+f_0}{1-f_2(e^{f_1}+f_0)},
\end{eqnarray}
and the inverse relation
\begin{eqnarray}
f_0&=&\frac {x_1x_2}{x_1-x_2} \cr
f_1&=&\log \left[\frac {x_1^2(x_3-x_2)} {(x_1-x_2)(x_1-x_3)}\right] \cr
f_2&=&-\frac {1} {x_1} .
\end{eqnarray}

We remark that now the third differential equation for $f_2$ 
in (\ref{nsistf}) 
is the same as in the case II) and it provides a new method of finding 
the general solution of (\ref{Riceq})  once the particular solution 
satisfying $f_2(0)=0$   of the associated Riccati equation is found.

\bigskip

V) The fifth possibility is

\be
U(t)=\exp(v_0L_0)\exp(v_2L_2)\exp(v_1L_1) ,
\ee
with associated system 
\begin{eqnarray}
\dot{v}_0&=&a_0e^{-v_1}\cr
\dot{v}_1&=&a_1-2a_0v_2e^{-v_1}\cr
\dot{v}_2&=&a_2e^{v_1}-a_0{v_2}^2e^{-v_1},\label{sysfifth}
\end{eqnarray}
and solution 
\be
x(t)=\frac {e^{v_1}(x_0+v_0)} {1-v_2(x_0+v_0)} ;
\ee
the expressions for the three particular solutions  $x_1, x_2, x_3 $  
are
\begin{eqnarray}
x_1(t)&=& -\frac {e^{v_1}}{v_2} \cr
x_2(t)&=&\frac {e^{v_1}v_0} {1-v_2v_0} \cr
x_3(t)&=& \frac {e^{v_1}(1+v_0)}{1-v_2(1+v_0)} ,
\end{eqnarray}
with the inverse relation 
\begin{eqnarray}
v_0&=& \frac {(x_3-x_1)x_2}{(x_2-x_3)x_1} \cr
v_1&=&\log \left[\frac {x_1^2(x_2-x_3)} {(x_2-x_1)(x_1-x_3)}\right] \cr
v_2&=&-\frac {(x_2-x_3)x_1} {(x_2-x_1)(x_3-x_1)} .
\end{eqnarray}

VI) The last possibility is the factorization 
\be
U(t)=\exp(w_2L_2)\exp(w_0L_0)\exp(w_1L_1),
\ee
and then the associated system is 
\begin{eqnarray}
\dot{w}_0&=&a_0e^{-w_1}-a_2w_0^2e^{w_1}\cr
\dot{w}_1&=&a_1+2a_2w_0e^{w_1}\cr
\dot{w}_2&=&a_2e^{w_1} ,\label{syssexth}
\end{eqnarray}
the general solution
\be
x(t)=\frac {e^{w_1}x_0} {1-w_2x_0} +w_0e^{w_1} ,
\ee
and the  expressions for  $x_1, x_2, x_3 $,  
\begin{eqnarray}
x_1(t)&=& -\frac {e^{w_1}}{w_2}+w_0e^{w_1} \cr
x_2(t)&=& w_0e^{w_1} \cr
x_3(t)&=& \frac {e^{w_1}}{1-w_2}+w_0e^{w_1} ,
\end{eqnarray}
with inverse relation 
\begin{eqnarray}
w_0&=& \frac {(x_3-x_1)x_2}{(x_2-x_1)(x_3-x_2)} \cr
w_1&=&\log \left[\frac {(x_2-x_1)(x_3-x_2)} {(x_3-x_1)}\right] \cr
w_2&=&\frac {(x_3-x_2)} {(x_3-x_1)} .
\end{eqnarray}

Finally, we remark that in these two last  approaches  there is no uncoupled 
differential equation of the  Riccati type whose solution allows us to find  
the solution of the two other 
remaining equations in the system, and therefore the general solution of 
the original  Riccati equation,
anymore. However, if in the third equation in (\ref{sysfifth}) we define
$\phi(t)=\dot v_1$ we will get the Riccati equation
\be
\dot \phi=\frac 12 \phi^2+q(t)\phi+p(t)
\ee
with
\be
q(t)= \frac{\dot a_2}{a_0},\quad p(t)=\dot a_1-2a_0a_2-\frac {a_1}{a_0}
\dot a_2+\frac{a_1^2}2.
\ee

In a similar way, in the sixth case, taking derivatives in the second equation
in (\ref{syssexth}) and after some manipulations, the equation for $\varphi=\dot w_1$
becomes a Riccati equation
\be
\dot \varphi=s(t)+r(t)\varphi+\frac 12\varphi^2
\ee
where
\be
r(t)=\frac{\dot a_2}{a_2},\quad s(t)=\dot a_1-
\frac{a_1}{a_2}\dot a_2+2a_0a_2-\frac 12 a_1^2.
\ee

Let us summarize the results we have got. We have reduced the problem
of finding
the general solution of the Riccati equation to the one of determining a
curve in $SL(2,\R)$
which is defined through its second class canonical coordinates, and this
leads to
a differential equation system. Once the curve in  $SL(2,\R)$ is known
we are able to find
the general solution of the Riccati equation. However, even if we are
 not able to find the  solution of the corresponding differential
equation system for the second class coordinates in the group, we know
the form (\ref{gral}) of the general solution of
the original Riccati equation. Even more,
given  a set of (three in the Riccati case) fundamental
particular solutions we can determine 
the function giving us the superposition principle (\ref{solgxp}).
 
We  have seen that 
the general solution of  Riccati equation is given by 
\Eq{
x(t)=\frac {x_0x_1(x_3-x_2)+x_2(x_1-x_3)}{x_0(x_3-x_2)+(x_1-x_3)},  \label{supf}
}
where  $x_0$ is a constant depending on the initial conditions.
We aim now to show how it is possible to reconstruct the original 
differential equation once the superposition formula is given. 
In the case of Riccati equation the superposition formula 
(\ref{supf}) is equivalent to 
$$
x_0=\frac {a-c\, x} {d\, x-b}
$$
and it is easy to check that taking derivatives in the 
preceding relation  we find 
$$
(\dot a-\dot c x-c\dot x)(d\, x-b)-(a-c\, x)(\dot d \,x+d\,\dot x-\dot b)=0
$$
from where we find the following expression for  $\dot x$ 
$$
\dot x=\frac {(\dot c d-c\dot d)} {(bc-ad)} x^2 
+\frac {(-\dot a d+a\dot d +\dot b c-b\dot c)} {(bc-ad)} x 
  +\frac {(\dot a b-a\dot b)} {(bc-ad)}, 
$$
that is a Riccati equation.

On the other side, if we assume that there is a superposition formula 
for Riccati equation and we try to determine the function $\phi$ giving  
that formula, $x=\phi (x_1,x_2,x_3,k)$, 
we will have 
$$
\dot x= \dot{\phi} = \pd {\phi} {x_1} \dot x_1+\pd {\phi} {x_2} \dot x_2
	   +\pd {\phi} {x_3} \dot x_3 =a_0+a_1\phi +a_2{\phi }^2
$$
from where the following system of partial differential equations is found
\begin{eqnarray}
&&\pd{\phi}{x_1}+\pd{\phi}{ x_2}
	   +\pd {\phi} {x_3} =1 \cr
&&x_1\pd{\phi} {x_1}+x_2\pd {\phi} { x_2}
	   +x_3\pd {\phi} {x_3} =\phi \cr
&&x_1^2\pd {\phi} {x_1}+x_2^2\pd {\phi} { x_2}
	   +x_3^2\pd {\phi} {x_3} ={\phi}^2  .
\end{eqnarray}

 Now a long computation leads to 
the following expression for  $\phi$:
\Eq{
x(t)=\frac {kx_1(x_3-x_2)+x_2(x_1-x_3)}{k(x_3-x_2)+(x_1-x_3)}.
}

\section{The superposition principle for first order differential
equation systems and Lie-Scheffers theorem}
\setcounter{equation}{0}

We are now interested in studying the existence of a superposition
principle for 
first order differential equation systems generalizing the one obtained
for the Riccati equation. More explicitly, given a system
 \be
\frac {dx^i}{dt}=X^i(x,t), \qquad i=1,\ldots,n,\label{nasys}
 \ee
 we ask whether  there exists a set $\{x^{(1)},\ldots,x^{(m)}\}$ of fundamental solutions and
 a function $\Phi:\R^{n(m+1)}\to \R^n$  such that the general solution of
 (\ref{nasys}) can be expressed as
$$x=\Phi(x^{(1)},\ldots,x^{(m)};a_1
,\ldots,a_n),$$ where $a_1,\ldots,a_n$ are constants related with the initial conditions.

Before studying the general case, we remark that we know that, at least in the case where the system is
the autonomous linear system
 $$\frac {dx}{dt}=Ax,\ \ x\in \R^n,$$
 with $A$ being a constant matrix, it defines a flow $\phi_t$ which can
 be considered as a curve on
 the general linear group  $GL(\R,n)$ given by $\phi_t(x)=e^{At}x$.
 The flow satisfies
 $$\frac {d\phi_t}{dt}=\phi_t\circ A$$
 and it can be determined  from a fundamental set of solutions, i.e.,
 when the $n\times n$ matrix $X$ whose columns are the vectors
 defining the solutions $\{x_1(t),\ldots, x_n(t)\}$,
 $$X(t)=(x_1(t),\ldots, x_n(t)) ,
 $$
 is an invertible matrix, then the equation 
 $$X(t)=e^{tA}X(0)$$
 shows that the evolution operator $e^{tA}$  is determined as
 $e^{tA}=X(t)X(0)^{-1}$, i.e., the fundamental system of solutions allows
 us to find the flow of our first order differential equation system.
 In other words, we have a possitive answer to our previous question with $m=n$ and
$\Phi $  being the linear map $\Phi (x^{(1)},\ldots,x^{(n)};a_1
,\ldots,a_n)=a_1\, x^{(1)}+\cdots+a_n\, x^{(n)}$.
 
 From here it is clear that
 with any linear differential equation system on $\R^n$ we can associate an
 equation on $GL(n,\R)$ by setting
 $$\dot g=gA,
 $$
 or
 $$g^{-1}\frac {dg}{dt}=A.
 $$
 The matrix  $A$ is an element of ${\goth{ gl}}(n,\R)$.
 
 Moreover, this way of finding the resolvent can also be used for
  time--dependent systems
 $$\frac {dx}{dt}=A(t)x.
 $$
  In this case if we have a fundamental set of solutions,  denoted by
$X(t)=(x_1(t),\ldots, x_n(t))$, from
  $$X(t)=R(t,0)X(0)
  $$
  we get $R(t,0)=X(t)X(0)^{-1}$.

We could equally well associate
  a time--dependent equation on  $GL(n,\R)$ by setting
 $$g^{-1}\frac {dg}{dt}=a^i(t)A_i,
 $$
 with $A_i$ being  elements of the natural basis of  ${\goth{ gl}}(n,\R)$.
 
 The point we want to remark is that if the
  evolution preserves some structure, then
 $$A(t)=\phi_t^{-1}\circ\frac {d\phi_t}{dt}$$
 lies in certain subalgebra of ${\goth{gl}}(\R,n)$. The flow now
 defines a one--parameter 
 family of transformations, and conversely, given a one--parameter
 family of transformations it will determine a vector field in $\R^n$,
  the corresponding fundamental vector field.
 
 The critical fact is that the general solution is determined by
   a linear operator from 
 a set of fundamental solutions, and this is a linear
 superposition principle. It is very
  natural to ask what happens when the vector field is nonlinear.
 The answer is that, at least 
  in some cases, to be explicited shortly, there is a kind of non--linear superposition principle,
 as it was proved by
 Lie (\cite{LS}). This nonlinear superposition principle is simply a
  generalization of the previous construction to those cases where the
 action of the group is not linear and $\R^n$ is replaced
for a manifold $M$.

In the general case of the system (\ref{nasys}), the Theorem for existence and
uniqueness of solutions of such systems tells us that there will be, for each
small enough $t$, a local diffeomorphism of $\R^n$ which establishes the correspondence
among the initial values and the corresponding ones for the explicit
value of the parameter $t$. In other words, the evolution is described by a
 curve $g_t$ in the group of diffeomorphisms of $\R^n$. We have seen that when
 we consider a linear autonomous system this curve lies in
the group $GL(n,\R)$ and is just the exponential of the matrix $A$ giving the system,
$g(t)=\exp tA$. In the linear time--dependent case, we also have a curve $g_t$
in $GL(n,\R)$ but it is not the exponential anymore: the only thing we
can say is that 
 $$A(t)=\phi_t^{-1}\circ\frac {d\phi_t}{dt}
$$
takes values in the Lie algebra ${\goth{gl}}(n,\R)$. Actually, the
solution is obtained by the Dyson, time--ordered, exponential.

The point is that for other types of vector fields the curve described by
$g_t$ belongs to other Lie subgroups of the group of
diffeomorphisms and these are just the cases for which
the idea of the nonlinear superposition principle is generalizable.
 
 The result established by Lie and Scheffers \cite{LS} is that
the general evolution defined by (\ref{nasys}) can be expressed in
terms of $m$ fundamental 
 solutions if there are $r$
 vector fieds $Y_1$, \ldots, $Y_r$, such that the vector field $X$,
 $$X=X^i(x,t)\pd{}{x^i}$$
  can be expressed as a linear combination 
 \Eq{X=a_1(t)Y_1+\cdots +a_r(t)Y_r\label{vfsp}} and furthermore
 the vector fields
 $$Y_\alpha=\xi^i_\alpha(x)\pd {}{x^i}$$
 close a finite dimensional real (or complex) Lie algebra, with dimension $r$, 
 i.e., there exist $r^3$  real numbers
 $c_{\alpha\beta}\, ^\gamma$ such that
 \Eq{[Y_\alpha,Y_\beta]=c_{\alpha\beta}\, ^\gamma\,  Y_\gamma.\label{Lalgcon}}
 Moreover, in this case $mn\geq r$. When $\xi^i_\alpha(x)
 =a^i_{\alpha j}x^j+b^i_\alpha$, with $a^i_{\alpha j}$ and $b^i_\alpha$
 arbitrary constants, the system is linear.

 Let us consider an effective action
of a Lie group $G$ of dimensi\'on $r$ on  $n$-dimensional  
differentiable manifold $M$,
$\Phi:G\times M\to M$, and $\Phi_g:M\to M$ and $\Phi_x:G\to M$ denote
the maps $\Phi_g(x)=\Phi_x(g)=\Phi(g,x)$, for $g\in G, \, x\in M$.
Choosing an initial point $x(0)$, every
curve $g:I\to G$ in the group determine a curve in
the manifold $M$ by
$$ x(t)=\Phi(g(t), x(0))=\Phi_{g(t)}( x(0))
=\Phi_{ x(0)}(g(t)),    
$$
and taking derivatives with respect to $t$ we see that the tangent
vectors to the curves
 $g(t)$ and $x(t)$, respectively, are related by 
$$\dot{ x}(t)=\Phi_{ x(0)*g(t)}\dot g(t). 
$$

Let us remark that $\dot g(t)\in T_{g(t)}G$ and  $\dot{x}(t)
\in T_{ x(t)}M$.

We can express   $\dot{   x}(t)$ in terms of   ${  x} (t)$: we recall that 
 if $  x_2=\Phi_{g}(  x_1)$, then
$$\Phi _{  x_2}= \Phi _{  x_1}\circ R_g,$$
where $R_g$ denotes right translation in the Lie group $G$,
because for any  $g'\in G$, 
$$\Phi _{  x_2}(g')=\Phi (g',  x_2)=\Phi(g', \Phi(g,  x_1))
= \Phi (g'g,  x_1)=(\Phi _{  x_1} \circ R_g)(g').
$$

Now, using the chain rule for computing the differentials we find that 
$$\Phi _{  x_2*} =\Phi _{  x_1*} \circ R_{g *},
$$
and then,
$$\Phi _{  x_1*} =\Phi _{  x_2*} \circ R_{g^{-1} *}.
$$

We can use this relation for $g=g(t)$ and $  x_1=  x(0)$ and then we
find the following expression for
 $\dot{  x} (t)$:   
$$
\dot{  x} (t) = \Phi _{  x (t)*e} (R_{g^{-1}(t)*g(t)}\dot g(t)). 
$$

Since $R_{g^{-1}(t)}$ is the right translation leading  $g(t)$ to
the neutral element $e\in G$ and 
$\dot g(t)\in T_{ g(t)}G$, then 
$R_{g^{-1}(t)*g(t)}\dot g(t)\in T_eG$  and  
we know that $ T_eG$ may be identified with the Lie algebra of  $G$,
$\goth g$. Moreover, for linear Lie groups, i.e., 
 subgroups of  $GL(n,\R)$, right translation reduces to
right multiplication by the corresponding  matrix, and hence
$R_{g^{-1}(t)*g(t)}\dot g(t)$ is just the product 
 $\dot g(t)\, g^{-1}(t)$.
 
 Let $\{e_\alpha\}_{\alpha=1}^r$ be   a basis
 of the corresponding Lie algebra $\goth g$ with defining relations
 $$[e_\alpha,e_\beta]=c_{\alpha\beta}\,^\gamma e_\gamma,
 $$
  and denote $X_\alpha$ the corresponding fundamental
 vector fields defined by $$(X_{\alpha}f)(m)=\frac d{dt}
\left[f(\exp(-te_\alpha)m)\right]_{|t=0},
 $$
 for any differentiable function $f$.
 We recall that in this case 
 $$[X_\alpha,X_\beta]=c_{\alpha\beta}\,^\gamma\, X_\gamma.
 $$
 
 Now, if the  time--dependent  vector field defining the system (\ref{nasys})  
 is of the form
 $$X(t)=a_\alpha(t) X_\alpha$$
 we associate with it the following differential equation on the Lie group 
 itself
 $$g^{-1}(t)\frac {dg}{dt}=a_\alpha(t) e_\alpha.
 $$
 
 In this way the given system is replaced by a higher dimensional system of
 first order linear equations. Or in other words, we have replaced the 
 original system of differential equations with a new system on the group $G$,
 but the 
 important point is that   it is enough to find a particular solution, 
 the one starting from the neutral element,  for obtaining  the general solution 
  of the system(\ref{nasys}): the solution starting from $x_0$ is given   
  by   $\Phi(g(t),x_0)$. 

  Moreover, even if we do not know the solution $g(t)$ of the new system,
  it is possible to find it from the knowledge of a convenient set of 
  particular solutions of (\ref{nasys}) such that (\ref{vfsp}) with 
  the additional condition (\ref{Lalgcon}). More explicitly, given 
  a curve $x_1(t)$ that is a particular solution of the given system,
  there are, in principle, different possible choices for the curve $g(t)$  
  such that  $x_1(t)=\Phi(g(t),x_1(0))$, because the stability group of
the point $x_1(0)$  
  may be nontrivial. If we choose a different particular solution, $x_2(t)$, then the 
  ambiguity reduces to the group intersection of the isotopy groups of   
   $x_1(0)$ and  $x_2(0)$. We can, if necessary iterate the procedure
until we 
   arrive to  a set of $m$ particular solutions   $x_1,\ldots,x_m$  allowing us 
   the determination of the curve   $g(t)$. Of course as we have $r$ unknown functions, the
   second  class canonical coordinates, and we have $mn$ conditions, it should be
$mn\geq r$. 
 
 More explicitly, a set of  $x_1,\ldots,x_m$  of solutions is said to be  a fundamental
  system of solutions, if
 \begin{eqnarray} x_1(t)&=&\Phi(g(t),x_1(0))\cr
 \ldots&=&\ldots\dots\cr
 x_m(t)&=&\Phi(g(t),x_m(0))
 \end{eqnarray} 
is a minimal set allowing us to solve for $g(t)$ via the
 implicit function Theorem. If this can be done we get
 $$g(t)=F(x_1(t),\ldots,x_m(t);x_1(0),\ldots, x_m(0)),
 $$
 and then any other solution can be written as 
 $$x(t)-\Phi(F(x_1(t),\ldots,x_m(t);x_1(0),\ldots, x_m(0)), x(0))=0.$$
 
 Therefore the left hand side of this relation defines a
  constant of the motion.   
 
 Starting with the action $\Phi :G\times M\to M$ we should find the
 minimal integer
 number      $m$ such that the isotopy group of the action of $G$
 on the product
 $M^m=M\times \cdots \times M$ ($m$ times), extended from $\Phi$ by  
 $\Phi^m(g,x_1,\ldots,x_m)=(\Phi(g,x_1),\ldots,\Phi(g,x_m))$, reduces
 to the neutral 
 element for a point such that any two coordinates are different.  
 
 We recall that the fundamental vector field corresponding to an element 
 of $\goth g$
 generating a one-parameter Lie subgroup contained in the isotopy group of a 
 point, vanishes in such a point, and conversely. Therefore, when 
 expressed in terms of fundamental vector fields
  that means that the extensions to $M^m$ of 
 the fundamental vector fields $X_\alpha$ do not vanish in a
a point whose  coordinates  are  different. The general solution then 
 is found by adding a new component and looking for constants of motion.
 
 The procedure is next illustrated with an example for
 the simplest case $n=1$. According to Lie's
 Theorem we should look for a finite dimensional real Lie algebra of
 differential operators
 $$X_\alpha=f_\alpha (x) \pd{}{x}.
 $$
 
 It can be shown that the only finite dimensional Lie algebra that
 can be found from vector fields in one real variable
 are $\goth{sl}(2,\R)$ and its subalgebras. The uniquely defined
 (up to a change of variables) realization of $\goth{sl}(2,\R)$
 is given by  
 $$X_0=x\pd{}x,\ X_-=\pd{}x,\ X_+=x^2\pd{}x.
 $$
 
 These vector fields close the $\goth{sl}(2,\R)$ Lie algebra 
 $$[X_0,X_-]=-X_-,\ [X_0,X_+]=X_+,\ [X_-,X_+]=2X_0.
 $$
 
 Therefore, it suffices to consider the case in which
 the time--dependent vector field $X$ defining the equation can be written as a
 linear combination
 $X=a_1X_0+a_2X_++a_0X_-$ with $a_0=a_0(t)$,  $a_1=a_1(t)$ and  $a_2=a_2(t)$
 real functions. For simplicity we will consider the case
 when $a_1$, $a_2$ and  $a_3$ are real numbers and then we obtain the
 differential equation
 $$\frac {dx}{dt}=X(x,t)=a_0+a_1x+a_2x^2,
 $$
 which is nothing but the well known
 Riccati equation.
 
 First we note that the determinant 
 $$\deter{ccc}{1&x_1&x_1^2\\1&x_2&x_2^2\\1&x_3&x_3^2}=
 (x_1-x_2)(x_2-x_3)(x_3-x_1)
 $$
 is generically different from zero, while the system
 $a+bx_1+cx_1^2=0, \
 a+bx_2+cx_2^2=0 $ always has a solution. Consequently, $m=3$
 in this case.
 
  Now, for obtaining the general solution we should define the
 vector fields 
 \begin{eqnarray}
 V_0&=&x\pd{}{x}+x_1\pd{}{x_1}+x_2\pd{}{x_2}+x_3
 \pd{}{x_3},\cr
 V_-&=& \pd{}{x}+ \pd{}{x_1}+ \pd{}{x_2}+ \pd{}{x_3},\cr
 V_+&=&x^2\pd{}{x}+x_1^2\pd{}{x_1}+x_2^2\pd{}{x_2}
 +x_3^2  \pd{}{x_3}
 \end{eqnarray}
 and look for a solution of the system 
 $$V_0f=V_+f=V_-f=0.
 $$
 
 This system of partial differential equations is integrable,
 because the vector fields $V_0$, $V_+$ and $V_-$ close  a Lie
  algebra, and therefore
 they define an integrable distribution.
 
 The last equation $V_-f=0$ tell us that the function $f$ depends
  only on the 
 differences $u_1=x-x_1$, $u_2=x_1-x_2$ and $u_3=x_2 -x_3$,
 because the 
 characteristic system is 
 $$\frac{dx}1=\frac{dx_1}1=\frac{dx_2}1=\frac{dx_3}1,
 $$
  and it   has as first integrals the differences $u_1=x-x_1$,
 $u_2=x_1-x_2$ and $u_3=x_2 -x_3$.
 Now, if $f(x,x_1,x_2,x_3)=\varphi (u_1,u_2,u_3)$,
 the condition $V_0f=0$ is written 
 $$u_1\pd{\varphi}{u_1}+u_2\pd{\varphi}{u_2}+u_3\pd{\varphi}{u_3}=0,$$
 i.e.,
 the function $\varphi$ should be homogeneous of degree zero, and
 therefore it only 
 can depend on the quotients $v_1=u_1/u_2$ and $v_2=u_3/u_2$,
 $\varphi(u_1,u_2,u_3)=\phi(v_1,v_2)$. Finally,
  the condition $V_+f=0$ can be written in these coordinates,
 after a long computation, as 
 $$v_1(v_1+1)\pd{\phi}{v_1}-v_2(v_2+1)\pd{\phi}{v_2}=0.
 $$
 
 The corresponding characteristic system is 
 $$\frac{dv_1}{v_1(v_1+1)}=-\frac{dv_2}{v_2(v_2+1)}
 $$ 
 and taking into account that 
 $$\int \frac{d\xi}{\xi(\xi+1)}=\log\frac \xi{\xi+1},$$
 we obtain that the constant of motion  $f$  should be a function of 
 $$\zeta= \frac{ v_1}{v_1+1}\frac{ v_2}{v_2+1},
 $$
 and therefore
 $$
 \frac{(x-x_1)(x_2-x_3)}{(x-x_2)(x_1-x_3)}=c
 $$
 provides the non--linear evolution principle giving $x(t)$
 as a function of three independent solutions
 $$x=\frac {(x_1-x_3)x_2c+x_1(x_3-x_2)}{(x_1-x_3)c
 +(x_3-x_2)}.
 $$
 
 Let us remark that for $n=1$ there is only one nonlinear
 differential equation family
 satisfying Lie--Scheffers  theorem: the Riccati equation. Of course,
 proper subalgebras of $\goth{sl}(2,\R)$ lead to the linear
 inhomogeneous equation when $a_2=0$ or to a
 linear homogeneous equation when $a_0=a_2=0$. However, for $n= 2$ in addition to
 $SL(3,\R)$, $O(3,1)$ and $O(2,2)$, we can realize
 families of Lie algebras with arbitrary large Abelian
 ideals.

\section{Application in the solution of second order
differential equations}
\setcounter{equation}{0}

Algebraic methods have very often been used in the search for 
eigenvalues of operators and the corresponding eigenvector. The
particular case of the harmonic oscillator is the prototype and 
it is based on creation and annihilation operators, and therefore 
it is related with the Heisenberg group. The possiblity of relating 
linear second order differential equations with  a Riccati equation, 
as indicated above, and the related group $SL(2,\R)$ has not 
been exploited  till now, as far as we know. 

In this section we will explore the use of Wei--Norman method based on 
the  $SL(2,\R)$ group for studying the spectral problem of the second order  
differential operator determined for the  Hamiltonian of 
the Harmonic oscillator
\be
H=\frac {P^2} {2M}+\frac {k} {2}X^2 ,
\ee
where  $k$ is a constant.

We will use the following notation:
\begin{eqnarray}
\w &=&\sqrt {k/M}, \\
\a &=&\sqrt {M\w/\hbar}, \\
\xi &=&\a x, \\
\lambda &=&2E/\hbar \w, 
\end{eqnarray}   
and then the  Hamiltonian can be written 
\be
H=\frac {\hbar \w} {2}\left[-\frac {d^2}{d{\xi}^2}+{\xi}^2\right].
\ee

The eigenvectors  $\psi (\xi)$ of the preceding Hamiltonian  operator
corresponding to the eigenvalues  $\lambda\frac{\hbar\omega}2$
are the  normalizable solutions of the 
 differencial    equation
\be
-\frac {d^2\psi}{d{\xi}^2}+{\xi}^2\psi=\la \psi.  \label{eqcar}
\ee

We proved in Section 2 that if  $ \psi$ is a solution of (\ref{eqcar}), 
then the function 
 $z=\frac{1}{\psi}\frac {d\psi}{d{\xi}}$ will be a solution of the 
 following Riccati  equation
\be
\frac{dz}{d\xi}=-z^2+\left({\xi}^2-\la \right). \label{Richar}
\ee
As it was stated in Section 3, such equation admits a nonlinear superposition 
principle based on the  $SL(2,\R)$ group, and therefore the general 
solution can be found by means of an appropriate factorization
\be
z(\xi)=\exp(g_2L_2)\exp(g_1L_1)\exp(g_0L_0)(z){\vert}_{\xi=0}.
\ee

The functions $g_0,g_1,g_2$, are to be determined from the first order
differential equation system 
\begin{eqnarray}
\dot g_0 &=& {\xi}^2-\la -g_0^2 \cr 
\dot g_1 &=& -2g_0 \\ \label{sysg}
\dot g_2&=& -e^{g_1},\nonumber
\end{eqnarray} 
together with the initial conditions $g_0(0)=g_1(0)=g_2(0)=0$.

Let us first remak that the Riccati equation 
$$
\f(dz,d\xi)+z^2-{\xi}^2+\la=0,
$$
under the change of variables given by
\be
z=2\xi v-\xi,\qquad 
y={\xi}^2,
\ee
becomes a new   Riccati equation, 
\be
\frac{dv}{dy}+v^2+v\left(\frac 1{2y}-1\right)-\frac{1-\la}{4y}=0,
\label{nuevaR}
\ee

On the other side, the Riccati equation associated, according to
 the method described in Section 2,
 with the linear second order
 confluent  hypergeometric differential equation
\Eq{
yW''+\left(b-y\right)W'-aW=0,\label{chde}
}
where  $a$  and   $b$  are constants and $W'$  and  $W''$ are the first
 and second
 derivative, respectively, of the function $W(y)$, is
\be
\frac{dv}{dy}+v^2+v\left(\f(b,y)-1\right)-\f(a,y)=0.\label{Ricashc}
\ee
with 
$$
v=\f(W',W).
$$  

 A simple comparison between (\ref{nuevaR}) and (\ref{Ricashc}) shows
 that both coincide when
 \be
a=\f(1-\la,4) \ \qquad
b=\f(1,2) .\label{parho}
\ee

It is well known that the general solution of (\ref{chde}) is given by 
$$
W(y)= AM\left(\f(1-\la,4),\f(1,2),y\right)+
		By^{\f(1,2)}M\left(\f(3-\la,4),\f(3,2),y\right)
$$
with $A$  and   $B$ arbitrary  constants, and  $M(a,b,y)$ is such that for
large values
of the variable $y$,
\begin{eqnarray}
M(a,b,y)&=&\f(\G(b),{\G(b-a)}) e^{i \epsilon \pi a}y^{-a}g(a,a-b+1,-y)  \\
&+&\f(\G(b),{\G(a)}) e^yy^{a-b}g(1-a,b-a,y),
\end{eqnarray}
where $\epsilon=1$ if $-\pi/2<\Arg y<3\pi/2$ and $\epsilon=-1$
when $-3\pi/2<\Arg y\leq -\pi/2$ and  $g$ denotes the  function
\be
g(a,b,y)=\sum_{n=0}^{\infty}\f({a(a+1)\cdots(a+n-1)b(b+1)\cdots(b+n-1)},
{n!\,y^n})
\ee
Therefore, the function  $W(y)$ behaves for large values of $y$ as
 $e^y=e^{\xi^2}$ unless that
\begin{eqnarray}
\G\left(\f(1,4)-\f(\la,4)\right)&=&\infty
\rm{\   and  \   }  B=0\\
 \G\left(\f(3,4)-\f(\la,4)\right)&=&\infty
\rm{\   and  \   }  A=0,
\end{eqnarray}

Now, we recall that 
Gamma function is such that $\Gamma(0)=-\infty $ and takes finite  values
when $-z$ is not a positive integer number, while  $\G(-m)=\infty$ for
$0<m\in {\Bbb Z}$. Therefore if we want to have a solution of (\ref{chde})
 with values $a$ and $b$ satisfying
(\ref{parho}) being square integrable, it is necessary that 
\begin{eqnarray}
\f(1,4)-\f(\la,4)&=-m,  {\rm{\   and\ then\  }} \la=2n+1
 {\rm  {\  with\   }}
n=2m=2,4,6,\cdots  \\
\f(3,4)-\f(\la,4)&=-m,  {\rm{\   and\ then\ }} \la=2n+1 {\rm  {\  with\  }}
n=2m+1=3,5,7,\cdots,  
\end{eqnarray}
or,
\begin{eqnarray}
\f(1,4)-\f(\la,4)&=0  {\rm{\   and\ then\ }} \la=2n+1 {\rm  {\  with\  }}
n=0  \\
\f(3,4)-\f(\la,4)&=0  {\rm{\   and\ then\ }} \la=2n+1 {\rm  {\  with\  }}
n=1 . 
\end{eqnarray}

Consequently, a  necessary condition for (\ref{chde})
 with values $a$ and $b$ satisfying
(\ref{parho}) to be square integrable is 
$$
\la=2n+1,
$$
where $n=0,1,2,3,\ldots\, $.

Therefore,  we will restrict ourselves  to the case 
  $\la=2n+1$. We will show that the solution for   $g_0$ 
  can be obtained by induction on the index $n$.

 We first consider the case in which 
  $n$ is an even number. 
 The first equation reduces for  $n=0$ to
\be
\dot g_0 = {\xi}^2-1 -g_0^2 ,
\ee
and then it is an easy matter to check that 
the solution we are looking for is  $g_0=-\xi$.

In the case  $n=2$  the equation becomes 
\be
\dot g_0 = {\xi}^2-5 -g_0^2 ,
\ee
and again it is easy to check that then the solution    is
\be
g_0=\frac{8\xi}{4{\xi}^2-2}-\xi.
\ee

 In a similar way, when studying the cases 
 $n=4,6,\ldots$ we will see that for an even number $ n$
 the solution satisfying  $g_0(0)=0$ is given by 
\be
g_0=\frac {H'_n(\xi)}{ H_n(\xi)} -\xi,
\ee
with  $H_n$ being the  Hermite polynomial of order  $n$ and  $H'_n$ means
the derivative of  $H_n$ with respect to $\xi$. In fact, using several properties of
 Hermite polynomials, we see that 
\begin{eqnarray}
\frac{H''_n}{H_n}-\frac {(H'_n)^2}{H_n^2}-1&=&{\xi}^2-\la 
-\left(\frac {H'_n}{H_n} -\xi \right)^2, \\
H''_nH_n-(H'_n)^2-H_n^2&=&\left({\xi }^2-\la \right) H_n^2-
\left(H'_n-H_n\xi \right)^2, 
\end{eqnarray}
and taking into account that $H'_n=2nH_{n-1}$ and the corresponding 
relation for the derivatives
$H''_n=4n(n-1)H_{n-2}$, we will get 
\begin{eqnarray}
4n(n-1)H_{n-2}H_n-H_n^2&=&-\la H_n^2+2\xi H_nH'_n ,\\
4n(n-1)H_{n-2}-H_n&=&-\la H_n+2\xi H'_n. 
\end{eqnarray}

Now, the recurrence relation $H_n=2\xi H_{n-1}-2(n-1)H_{n-2}$, 
leads to
\be
2(n-1)H_{n-2}\left(2n+1-\la \right) =2\xi H_{n-1}\left(2n+1-\la \right),
\ee
and the right hand side vanishes because of  $\la =2n+1$.
 
In summary, we have checked that for any even number   $n$ 
the solution such that 
 $g_0(0)=0$ is   given by 
\be
g_0=\frac{H'_n}{H_n}-\xi.
\ee

When introducing this value for $g_0$ in the second equation (\ref{sysg})
for  $g_1$ we obtain the new equation 
\be
\dot g_1=-2g_0=-2\left(\f (H'_n,H_n)-\xi \right),
\ee
which can be easily integrated 
\be
g_1=-2\int \left(\frac{H'_n}{H_n}-\xi \right)d\xi =
-2\log H_n+{\xi}^2+C_1 =-\log \left(H_n^2+e^{{\xi}^2} \right) +C_1,
\ee
and as  $g_1(0)=0$, $C_1$ takes the value  
$C_1=\log \left(H_n^2(0) \right)=\log k_1$.
From the relation $H_n=2\xi H_{n-1}-2(n-1)H_{n-2}$ we see that 
$H_n(0)=-2(n-1)H_{n-2}(0)$ and iterating this reasoning we obtain 
the chain of relations 
\begin{eqnarray}
H_{n-2}(0)&=&-2(n-3)H_{n-4}(0)\\
H_{n-4}(0)&=&-2(n-5)H_{n-6}(0)\\
\cdots & =& \cdots \cdots  \\
H_{2}(0)&=&-2(n-n+1)H_0(0)=-2\, \cdot\,1
\end{eqnarray}
and therefore,
\begin{eqnarray}
H_n(0)&=&\left(-2(n-1)\right)\left(-2(n-3)\right)\left(-2(n-5)\right)
		\cdots \left(-2.1\right) \cr
	&=&\left(-2\right)^{n/2}\left(n-3\right)\left(n-5\right)
		\cdots 1, 
	\end{eqnarray}
and then,
\begin{eqnarray}
k_1=H_n^2(0)&=&\left[2^{n/2}(n-1)(n-3)(n-5)\cdots 1\right]^2 \cr
	&=&\left[2^{n/2}\frac{{n(n-1)(n-2)(n-3)\cdots 1}}{{n(n-2)(n-4)
\cdots 2}}\right]^2 \cr
&=&\left[\frac{n!}{{\frac n2\left(\frac n2-1\right)\left(\frac n2-2\right)
\cdots 1}}\right]^2 
=\left[\frac{n!}{{\left(\frac n2\right)!}},\right]^2.
\end{eqnarray}

 So, the function $g_1$ is given by the  expression
\be
g_1=-\log\left[k_1H_n^2e^{-\xi^2}\right].
\ee

Finally, the function  $g_2$ is to be determined from the
differential equation
\be
\dot g_2=-\f (1,{k_1H_n^2e^{-\xi^2}}),
\ee
whose  solution is 
\be
g_2=- \int \f (1,{k_1H_n^2e^{-\xi^2}})d\xi +k_2 
\ee
where  $k_2$ is to be chosen such that   $g_2(0)=0$, i.e.,
\be
k_2=\left[\int \f (1,{k_1H_n^2e^{-\xi^2}})\right]_{(\xi =0)}.
\ee
 
Putting together all previous results we get for the general solution 
 $z(\xi)$ the following expression:

\begin{eqnarray}
z(\xi) &=&\frac{{\left[e^{g_1}z\right]_{(\xi=0)}}}{{1-\left[g_2z
\right]_{(\xi=0)}}}     +g_0 \cr
&=&\frac{{\left[\frac1{{k_1H_n^2e^{-\xi^2}}}z\right]_{(\xi=0)}}}{
{1+\left[\left[\int \frac {1}{{k_1H_n^2e^{-\xi^2}}}\,d\xi -k_2 \right]z
\right]_{(\xi=0)}}}
+\frac{H'_n}{H_n} -\xi \cr
&=&\left[\frac {{\frac{1}{{k_1H_n^2e^{-\xi^2}}}z}}{
{1+\left[\int \frac{1}{{k_1H_n^2e^{-\xi^2}}}\, d\xi -k_2 \right]z}}
\right]_{(\xi=0)}+\frac{H'_n}{H_n} -\xi.        \label {solgn}
\end{eqnarray}
 
 We recall that the wave function $\psi$ was given by 
\be
\psi=e^{\int z(\xi)d\xi},
\ee
and from  
\be
\int z(\xi)d\xi=\log H_n-\frac{\xi^2}{2},
\ee
we obtain
\be
\psi=e^{\log H_n-\frac{\xi^2}{2}}=H_n e^{-\frac{\xi^2}{2}}.     \label {solHO}
\ee
\bigskip

 Notice that when $\psi $ has a well defined parity, the quotient 
 $z=\frac{1}{\psi}\frac {d\psi}{d{\xi}}$  is an odd function and 
 then the limit when  $\xi\to 0$ of $z $ should be zero. 
 However, when $\psi $ is a continuous   odd function, then 
 $ \lim_{\xi\to 0}\psi=0$ and then the quotient  
 $\frac{1}{\psi}\frac {d\psi}{d{\xi}}$ cannot be finite. This was the 
 reason for leaving aside the case $\lambda=2n+1$ for an odd number $n$.
 In this last case it is convenient first to introduce the change of variable 
$z=1/v$, and then the equation (\ref{Richar}) becomes 
\be
\frac{dv}{d\xi}=1-\left({\xi}^2-\la \right)v^2,
\ee
and the corresponding system of differential equations 
\begin{eqnarray}
\dot g_0 &= 1-\left({\xi}^2-\la \right)g_0^2 \cr
\dot g_1 &= -2\left({\xi}^2-\la \right)g_0 \\
\dot g_2 &= -\left({\xi}^2-\la \right)e^{g_1}. \nonumber
\end{eqnarray}  
with initial conditions $g_0(0)=g_1(0)=g_2(0)=0$.

It is now easy to check that the solution for $g_0$ is   
\be
g_0=\frac{H_n}{H'_n-\xi H_n},
\ee
that $g_1$ is given by 
\be
g_1={\xi}^2+\log\left[\frac{k_1}{H'_n-\xi H_n}\right]^2
\ee
and   that $g_2$  is
\be
g_2=-\int \frac{{\left({\xi}^2-\la \right)k_1}}{{\left(H'_n-\xi H_n\right)^2
e^{-{\xi}^2}}}\, d\xi +k_2,
\ee
where 
\begin{eqnarray}
k_1&=&\frac{2n!}{{\left(\frac{n-1}2\right)!}}, \\
k_2&=&\left[  \int \frac{{\left({\xi}^2-\la \right)k_1}}{
{\left(H'_n-\xi H_n\right)^2e^{-{\xi}^2}}}\, d\xi   \right]_{(\xi=0)}
\end{eqnarray}   
 
Then, under the change    $z=1/v$   the general solution 
\be
v(\xi)=\frac{{\left(e^{g_1}v\right)_{(\xi=0)}}}
{1-\left(g_2v\right)_{(\xi=0)}} +g_0,
\ee
becomes after some computations the expresion given by (\ref{solgn}).

\section{Conclusions}
The analysis developed in this paper
of the reduction process leading to the
Riccati equation starting from a second order linear differential equation 
just by a simple application of the general Lie theory of symmetry of
differential
 equations, now a well established subject  (see e.eg. \cite{Ol}), allows
us a better understanding  of the so called factorization method,
which is a
method
 that have been used for finding new solvable
potentials once one of such problem is known and that motivated
the study of supersymmetric quantum mechanics. The Riccati equation
was chosen here as the simplest example of first order
differential equation systems  admitting a nonlinear superposition
 principle and the deep reasons for the existence of such principle
 have been clarified in this paper. We hope that this will allow a
geometric interpretation of the converse part of Lie-Scheffers theorem.
This nonlinear superposition principle is very important and has played
a very important role because it provides an explicit expression for
 the solution of nonlinear equations admitting such superposition
principle. For a review see e.g. the review lecture
given by Winternitz \cite{PW}. We have also developed many examples
of the use of Wei--Norman method of computation of the solution of
differential equations in a Lie group  and we have applied it as
an academical example for reobtaining the eigenvalues and eigenvectors
of the harmonic oscillator problem.

\end{document}